\documentclass[epsfig,12pt]{article}
\usepackage{epsfig}
\usepackage{graphicx}
\usepackage{array}

\newcommand{\beq}{\begin{equation}}   
\newcommand{\eeq}{\end{equation}}
\newcommand{\beqn}{\begin{eqnarray}}   
\newcommand{\eeqn}{\end{eqnarray}}

\newcommand{\gsim}{\lower.7ex\hbox{$
\;\stackrel{\textstyle>}{\sim}\;$}}
\newcommand{\lsim}{\lower.7ex\hbox{$
\;\stackrel{\textstyle<}{\sim}\;$}}
\begin{document}

\begin{titlepage}

\begin{flushright}
FTPI-MINN-13/17, UMN-TH-3205/13\\
May 29/2013
\end{flushright}

\vspace{0.7cm}

\begin{center}
{  \large \bf  Calculating Extra (Quasi)Moduli  on the \\ [0.8mm] Abrikosov-Nielsen-Olesen string with 
\\ [2mm]
Spin-Orbit Interaction}
\end{center}
\vspace{0.6cm}

\begin{center}
 {\large 
  Sergei Monin,$^{a}$   M. Shifman,$^b$ A. Yung$^{b,c}$}

\end {center}

 \vspace{3mm}
 
\begin{center}
$^a${\em Department of Physics, University of Minnesota,
Minneapolis, MN 55455, USA} \\[1mm]
$^b${\em William I. Fine Theoretical Physics Institute, University of Minnesota,
Minneapolis, MN 55455, USA} \\[1mm]
$^{c}${\em Petersburg Nuclear Physics Institute, Gatchina, St. Petersburg
188300, Russia
}

\end {center}

\vspace{2cm}

\begin{center}
{\large\bf Abstract}
\end{center}

Using a representative set of parameters we
 numerically calculate the low-energy Lagrangian on the world sheet of the Abrikosov-Nielsen-Olesen string
in a model  in which it acquires rotational (quasi)moduli. The bulk model is deformed by a
spin-orbit interaction generating a number of ``entangled" terms on the string world sheet.

\hspace{0.3cm}

\end{titlepage}

\newpage

\section{Introduction}

In the previous publications simple models with
``spin-orbit" interactions supporting the Abrikosov-Nielsen-Olesen (ANO) \cite{ANO} or similar strings (vortices) 
were considered \cite{A,B,C} with ``extra" non-Abelian moduli (or quasimoduli) on the string world sheet.
Such extra moduli fields can appear in the bulk models  that have order  parameters carrying spatial indices, such as 
those relevant for superfluidity in $^3$He (see e.g. \cite{5}).  This particular example was studied in  \cite{B},
which, in fact inspired a more detailed numerical analysis presented below. The  studies in \cite{A,B,C} were carried out at a qualitative level.
Here we perform calculations needed for the proof of stability of the relevant solutions and derivation
of all constants appearing in the low-energy theory on the
string world sheet.

First, we will consider the simplest
model \cite{A} assuming weak  coupling in the bulk (to justify the quasiclassical approximation), determine the profile functions to find the string solution,  and derive the world sheet model. The general theory of the string moduli in the absence of the spin-orbit terms is discussed in \cite{IO,3}.

Then we introduce a spin-orbit interaction in the bulk. The impact of this interaction on the string (vortex) world sheet
amounts to lifting all or some rotational zero modes (i.e. those not associated with the spontaneous breaking of the translational symmetry by the string). However, under certain condition on a parameter determining the
spin-orbit interaction in the bulk, the mass gap generated on the world sheet remains small, and the extra zero modes survive as quasizero modes (some may remain at zero  at the classical level). 
In addition to the above mode-lifting, the spin-orbit interaction 
generates a number of interesting entangled terms on the string world sheet which couple rotational and translational modes (despite the fact that the translational modes remain exactly gapless).

\section{Formulation of the problem}
\vspace{2mm}

We start from the model suggested in \cite{A}. Its overall features are similar to those of the 
superconducting cosmic strings \cite{Wi}.
The model is described by an effective 
Lagrangian
\beq
{\cal L} = {\cal L}_{0} +{\cal L}_\chi
\eeq
where
 \beqn
{\cal L}_{0} &=& -\frac{1}{4e^2}F_{\mu\nu}^2 + \left| {\mathcal D}^\mu\phi\right|^2
  -V(\phi )\, ,
  \nonumber\\[2mm]
  {\mathcal D}_\mu\phi &=& (\partial_\mu -iA_\mu )\phi\,,
\nonumber\\[2mm]
  V &=& \lambda \left(|\phi |^2 -v^2
\right)^2\,,
  \label{tpi16}
\eeqn
and
\beqn
{\cal L}_\chi &=& \partial_\mu \chi^i \, \partial^\mu \chi^i  - U(\chi, \phi)\,,
\label{14}\\[2mm]
U &=&  \gamma\left[\left(-\mu^2 +|\phi |^2
\right)\chi^i \chi^i + \beta \left( \chi^i \chi^i\right)^2\right],
\label{15p}
\eeqn
with self-evident definitions of the fields involved, the covariant derivative, and the kinetic and potential terms.
The parameters  $e$, $\lambda$, $\beta$, $\mu$, and $v$ can be chosen at will, with some mild constraints
(e.g. $v>\mu$) discussed in \cite{B}. 
In particular, the stability of the $\phi\neq 0$ vacuum we are interested in implies that $\beta$ cannot be too small,
\beq
 \beta \geq \frac{m_\chi^2}{m_\phi^2}\,\frac{1}{c(c-1)}\,,
\eeq
where
\beq
c\equiv \frac{v^2}{\mu^2}\,,
\eeq
cf. Eq. (\ref{h8}).
The relations between the parameters in (\ref{tpi16}), (\ref{15p}) and $a,b,c$ appearing below, on the one hand,  and the physical parameters (the particle masses and the coefficients in front of the quartic terms $\phi^4$, $\chi^4$ and $\phi^2\chi^2$, respectively), on the other hand,  are shown in Table \ref{t1} and (\ref{h6}), (\ref{h8}). 

\vspace{0.2cm}
\newcolumntype{C}{ >{\centering\arraybackslash} p{3cm} }
\newcolumntype{D}{ >{\centering\arraybackslash} p{3cm} }
\begin{table}[h]
\centering
\renewcommand\arraystretch{3.5}
\begin{tabular}{|C|D|}
\hline
$\beta$ & $\displaystyle\frac{\tilde\lambda}{\gamma}$ \\\hline
$a$ & $\displaystyle\frac{m_A^2}{m_{\phi}^2}$ \\\hline
$b$ & $\displaystyle\frac{m_{\chi}^2}{m_{\phi}^2}$ \\\hline
$\displaystyle\frac{v^2}{\mu^2}\equiv c$ & $\displaystyle \left( 1-\frac{4\lambda}{\gamma}\frac{m_{\chi}^2}{m_{\phi}^2}\right)^{-1}$ \\\hline
\end{tabular}
\caption{\small Parameters in (\ref{tpi16}), (\ref{15p}) in terms of the particle masses and the coefficients in front of 
the quartic terms $\phi^4$, $\chi^4$, and $\phi^2\chi^2$  ($\lambda$, $\tilde\lambda$, and $\gamma$, respectively).}
\label{t1}
\end{table}

We will assume the parameters  to be chosen in such a way that
the bulk model is weakly coupled and, hence, the quasiclassical approximation is applicable.

Now let us discuss some parameters and the corresponding notation. In the vacuum the complex field $\phi$ develops a vacuum expectation value $|\phi_{\rm vac} | = v$ while its phase is eaten up by the Higgs mechanism. The masses of the (Higgsed) photon and the Higgs excitation are
\beq
m^2_A=2e^2v^2\,,\quad m^2_{\phi}=4\lambda v^2\, .
\label{h6} 
\eeq
We will denote the ratio of the masses
\beq
a=m^2_A/m^2_{\phi}\equiv \frac{e^2}{2\lambda}\,\,. 
\eeq
Moreover, in the vacuum the field  $\chi^i$ does {\em not} condense. Its mass is
\beq
m^2_{\chi}=\gamma\left(v^2-\mu^2\right).
\label{h8}
\eeq
For what follows we will introduce two extra dimensionless parameters:
\beq
b=m^2_{\chi}/m^2_\phi \equiv \frac{\gamma}{4\lambda}\,\frac{c-1}{c}\,,\quad c=v^2/\mu^2\,.
\eeq
The first measures the ratio of the $\chi$ to $\phi$ masses in the bulk and, as explained in \cite{A}, has to be $b\gsim 1$. 
The second parameter is also constrained, $c>1$. We will treat both of them as parameters of the order of unity.
As for the spatial orientation, the string will be assumed to lie along the $z$ axis. We introduce a dimensionless radius in the perpendicular $\{x,y\}$ plane,
\beq
\rho = m_\phi \, \sqrt {x^2 + y^2 }\,.
\eeq

The basis of our construction is the standard ANO string (see e.g. \cite{8}). The $\phi$ field winds ensuring topological stability, which entails in turn  its  vanishing at the origin. This implies the
following 
ans\"{a}tze:
\beq
 A_0=0\,,\quad A_i=-\varepsilon_{ij}\frac{x_j}{r^2}\left(\rule{0mm}{4mm} 1-f(r)\right)\,,\quad
\phi=v\varphi(\rho )e^{i\alpha}\,,
\label{aaa}
\eeq
where $\alpha$ is the polar angle in the perpendicular plane, and we assume for simplicity the minimal (unit) winding.
The boundary conditions supplementing (\ref{aaa}) are
\beq
f(\infty)=0\,,\qquad f(0)=1\,;\qquad\,
 \varphi(\infty)=1 \,,\quad \varphi(0)=0\,.
 \label{bc1}
\eeq

In the core of such a tube the $\phi$ field tends to zero, see (\ref{bc1}). 
The vanishing of the $\phi$ field results in
 the $\chi^i$ field  destabilization in the core of the string (as follows from Eq.~(\ref{15p})). 
Hence, inside the core, the $\chi^i$ field no longer vanishes,
\beq
(\chi^i\chi^i)_{\rm core} \approx \frac{\mu^2}{2\beta}\,,
\label{core}
\eeq
as will be illustrated by the graphs given below. Choosing the value of $\lambda$ judiciously, we
can make $\mu^2/\beta \gg m_\chi^2$, implying that the O(3) symmetry is broken in the core.
The appropriate ansatz is
\beq
\chi^i =\frac{\mu}{\sqrt{2\beta}}\,\chi(\rho)\left(\begin{array}{c}0\\0\\1\end{array}\right)\,,
\label{13p}
\eeq
with the boundary conditions
\beq
\chi(\infty)=0\,,\quad  \chi(0)\approx1\,.
\eeq
Thus, we have three profile functions, $f$, $\varphi$, and $\chi$, depending on $\rho$. Minimizing the energy functional we
derive the system of equations for the profile functions
\beqn
\left(\frac{f^{\prime}}{\rho}\right)^\prime&=&a\frac{\varphi^2\,f}{\rho} \,,\nonumber\\[2mm]
\left( \rule{0mm}{4mm} \phi^{\prime}\rho\right)^\prime &=&
\frac{f^2\,\varphi}{\rho}+\frac{\rho\varphi\left(\varphi^2-1\right)}{2}+\frac{\rho\,\varphi\,\chi^2}{2\beta}
\frac{b}{c-1}\,,
\nonumber\\[2mm]
\left(\rule{0mm}{4mm} \chi^{\prime}\rho\right)^\prime
&=& \frac{b}{c-1}\rho\chi\left(c\varphi^2+\chi^2-1\right)\,,
\label{15}
\eeqn
where the primes denote differentiation with respect to $\rho$.
In the numerical solution to be presented below we will assume for simplicity that 
\beq
 a=1\,,\,\,\,\mbox{ i.e.} \,\,\, m_\phi = m_A\,.
 \eeq
  In the absence of the $\chi$ filed
this would imply the Bogomol'nyi-Prasad-Sommerfield (BPS) limit \cite{BPS} with the tension\,\footnote{Alternatively, this is the boundary between type-I and type-II superconductors.}
\beq
T_0 =2\pi v^2\,.
\eeq
Below we will see how the presence of the $\chi$ filed changes the tension, using $T_0$ as a reference point.

It is obvious that the solution $\chi=0$ and $\varphi = \varphi_0\equiv \varphi_{\rm ANO}$ satisfies the
set of equations (\ref{15}). First we will show that this solution is unstable, i.e. corresponds to the maximum rather than minimum of the energy functional.

\section{Instability of the \boldmath{$\chi=0$} solution}
\label{inst}

To prove instability we must demonstrate that for  $\varphi = \varphi_0\equiv \varphi_{\rm ANO}$
there is a negative mode in $\chi$, in much the same way as in \cite{Wi}.
To this end it is sufficient to examine the energy functional in the quadratic in $\chi$ approximation,
\beq
{\mathcal E}_\chi = \frac{\mu^2}{2\beta}\, L\, \int dx\,dy \left\{\chi\left[-\Delta +\gamma\mu^2\left(
-1+ \frac{v^2}{\mu^2}\varphi^2_0\right)\right]\chi\right\},
\eeq
where $L$ is the string length (tending to infinity),
and find the lowest eigenvalue of
\beq
\left[-\Delta+\gamma\mu^2\left(
-1+ \frac{v^2}{\mu^2}\varphi^2_0\right)\right]\chi =E \chi\,.
\label{h20}
\eeq
One can view (\ref{h20}) as a two-dimensional Schr\"odinger equation. Given that the ground state is spherically symmetric
and introducing
\beq
\psi (\rho )=\chi \,\sqrt{\rho}\,,
\eeq
one can rewrite (\ref{h20}) as
\beq
-\psi^{\prime\prime}+
\left( b\,\frac{c\varphi^2_0-1}{c-1}-\frac{1}{4\rho^2}\right)\psi =\epsilon\psi\,,\qquad \epsilon = 
\frac{E}{m_\phi^2}\,,
\eeq
where prime denotes differentiation over $\rho$.
Numerical solution at $c=1.25$ yields
\beq
\epsilon = \left\{
\begin{array}{c}
-1.479 \,\,\,\mbox{at}\,\,\, b = 1\,,\\[1mm]
-4.19 \,\,\,\mbox{at}\,\,\, b = 2\,.
\end{array}
\right.
\eeq

\section{\boldmath{$\chi\neq 0$} solution} 

To find the asymptotic behavior of the profile functions at $\rho\rightarrow\infty$ one can linearize these equations in this limit,
\beq
 f\sim\sqrt{\rho}\,e^{-\rho}\,, \quad \left(1-\varphi\right) \sim\frac{1}{\sqrt{\rho}}\,e^{-\rho}\,,\quad
 \chi\sim\frac{1}{\sqrt{\rho}}\, e^{-\sqrt{b}\rho}\,.
\eeq

We integrated Eqs. (\ref{15}) numerically for a number of points in the  parameter space $\{b,c,\beta\}$ keeping
$a=1$. Then the parameter $\lambda$ appears only as an overall factor, with the analytically known  dependence.
Representative plots are given in Figures 1 and 2 (at the end). The first plot at the very top is given to show the domain of $\rho$ in which an ``effective"
$m^2$ for the $\chi$ field is negative forcing $\chi^i$ to condense in the core. 
This is the domain
of negative $\chi^i$ contribution to the potential energy. 
Then the three profile functions are
presented: $f(\rho )$, $\varphi (\rho)$, and $\chi (\rho)$ (from top to bottom).
In terms of the physical parameters, 
Figure 1 corresponds to $m_\chi^2 = m_\phi^2$ and $\tilde{\lambda} = 160\lambda$ while Figure 2 corresponds
to $m_\chi^2 = 2m_\phi^2$ and $\tilde{\lambda} = 640\lambda$.

These plots demonstrate that $\chi(0)$
is indeed close to unity. In scanning the parameter space we observe  that
(i) increasing the parameter  $b$  (i.e. the $\chi$ mass)  increases both the width of the domain where the ``effective"
$m^2$ for the $\chi$ field is negative and the  value of $\chi(0)$, but  decreases the tension of the string;
(ii) increasing the parameter  $c$ (i.e. decreasing $\mu$)  acts in the opposite direction; (iii)
increasing the parameter $\beta$ acts in the same way as increasing $c$  but with a weaker impact.

\section{The world-sheet theory without spin-orbit term}

Now let us introduce moduli. Two translational moduli are obvious. Since they are well studied we will
not dwell on this part. Of interest are the rotational moduli. Given the nontrivial solution (\ref{13p})
we can immediately generate a family of solutions which go through the system of equations 
(\ref{15}), namely,
\beq
\chi^i =\frac{\mu}{\sqrt{2\beta}}\,\chi(\rho)\, S^i\,,
\label{13}
\eeq
where  the moduli   $S^i $ are constrained ($i=1,2,3$),
\beq
S^i\,S^i =1\,,\;
\eeq
therefore, in fact, we have two moduli, as was expected.
To derive the theory on the string world sheet we, as usual, introduce $t$, $z$ dependence converting the $S^i$ moduli into the moduli fields $S^i(t,z)$, and
\beq
\chi^i =\frac{\mu}{\sqrt{2\beta}}\,\chi(\rho)\, S^i(t,z)\,.
\eeq
Substituting this in the Lagrangian (3) and (4) we obtain the low-energy effective action
\beq
S=\frac{1}{2g^2}\int dt\,dz\,\left(\partial_kS^i\right)^2\,,\quad  k=t,z.
\eeq
where
\beq
\frac{1}{2g^2}=\frac{1}{8c\beta\lambda}\int^{\infty}_0 2\pi\rho\,\chi^2(\rho)\,d\rho\,.
\eeq
One can rewrite this as
\beq
\frac{g^2}{2\pi}=\lambda\, \frac{\beta}{\pi^2}\,\frac{c}{I_1}\,,
\eeq
where 
\beq
I_1=\int^{\infty}_0 \rho\,\chi^2(\rho)\,d\rho\,.
\eeq
For the parameters we used in Figs.$\,$1,$\,$2 we obtain
\beq
I_1\approx 1.107\;\;\mbox{(for  Fig. 1)}, \qquad I_1\approx 1.18\;\;\mbox{(for  Fig. 2)}\,,
\eeq
and, correspondingly,
\beq
\frac{g^2}{2\pi}\approx 0.915\,\lambda\;\;\mbox{(for  Fig. 1)}\,, \qquad \frac{g^2}{2\pi}\approx 
1.717\, \lambda\;\;\mbox{(for  Fig. 2)}\,.
\eeq

\section{Spin-orbit interaction}

The ``two-component"  $\phi$-$\chi$  string solution presented above spontaneously breaks two translational symmetries, in the perpendicular $x,y$ plane, and O(3) rotations. The latter are spontaneously broken by the string orientation along the $z$ axis (more exactly, O(3)$\to $O(2)), and by the orientation of the spin field $\chi^i$ inside the core of the flux tube introduced through $S^i$. 

Now, we deform Eq. (\ref{14}) by adding a  spin-orbit  interaction \cite{C},
\beq
{\cal L}_\chi = \partial_\mu \chi^i \, \partial^\mu \chi^i -\varepsilon (\partial_i\chi^i)^2 - U(\chi, \phi)\,,
\label{28}
\eeq
where $\varepsilon$ is to be treated as a perturbation parameter.

If $\varepsilon = 0$ (i.e. Eq.~(\ref{14}) is valid)  the breaking O(3)$\to $O(2)  produces no extra zero modes 
(other than translational) in the $\phi$-$A_\mu$ 
sector \cite{IO,3}.
Due to the fact that $\chi\neq 0$ in the core, we obtain
 two extra moduli $S^i$ on the world sheet. This is due to the fact that at $\varepsilon = 0$
the rotational O(3) symmetry is enhanced \cite{B,C} because of the  O(3)
rotations of the ``spin" field $\chi^i$, independent of  the coordinate spacial rotations. 

What happens at $\varepsilon\neq 0\,$, see Eq.~(\ref{28})? If 
$\varepsilon$ is small, to the leading order in this parameter, we can determine the effective world-sheet action using the
solution found above at $\varepsilon = 0$. Two distinct O(3) rotations mentioned above become entangled: 
O(3)$\times$O(3) is no longer the exact symmetry of the model, but, rather, an approximate symmetry.
The low-energy effective action on the string world sheet takes the form
 \beqn
 S&= &\int dt\,dz\left(  {\cal L}_{{\rm O}(3)} + {\cal L}_{x_\perp} \right),
  \nonumber\\[2mm]
{\cal L}_{{\rm O}(3)} &=& \left\{ \frac{1}{2g^2} \left[\left(\partial_k S^i\right)^2  - \varepsilon \left(\partial_zS^3\right)^2\right]
 \right\}
 -
 M^2 \left(1- (S^3)^2\right),
   \label{five}\\[2mm]
{\cal L}_{x_\perp}&=& \frac{T}{2} \left(\partial_k \vec x_\perp\right)^2-M^2(S^3)^2\left(\partial_z\vec{x}_{\perp}\right)^2
\nonumber\\[2mm]
&+&
2M^2
\left( S^{3}\right) \left(S^1\partial_z x_{1\perp}+S^2\partial_z x_{2\perp}\right),
 \label{six}
 \eeqn
where $\vec x_\perp =\{ x(t,z),\, y(t,z)\}$ are the translational moduli fields,
 and $T$ is the string tension. The mass term $M^2$ is
\beq
 M^2 =  {\varepsilon} \,v^2 \,\frac{\pi I_2}{2c\beta}\,,
\eeq
where 
\beq
I_2=\int^{\infty}_0\rho\,(\chi^{\prime}(\rho))^2\,d\rho\,.
\eeq
For the values of parameters  used in Figs.$\,$1,$\,$2 we obtain
\beq
I_2\approx 0.378\;\;\mbox{(for  Fig. 1)}, \qquad I_2\approx 0.467\;\;\mbox{(for  Fig. 2)}\,.
\eeq
As for the tension $T$ we have
\beq
\frac{T}{T_0} \approx 0.963 \;\;\mbox{(for  Fig. 1)}\,,\qquad \frac{T}{T_0} \approx 0.953 \;\;\mbox{(for  Fig. 2)}\,.
\eeq
The impact of the $\chi^i$ field on the string tension is rather small and negative. The positive contribution of its kinetic energy is compensated by the negative potential energy, see Figs. 1 and 2 at the end. This was expected given the result of
Sec. \ref{inst}.

Moreover, it is seen that $$\frac{M^2}{T} \sim \frac{\varepsilon}{\beta} $$ and is small for sufficiently small ratio $\varepsilon/\beta $.
This justifies the above calculation.

\begin{figure}
 \begin{minipage}{7cm}
\includegraphics[width=7cm]{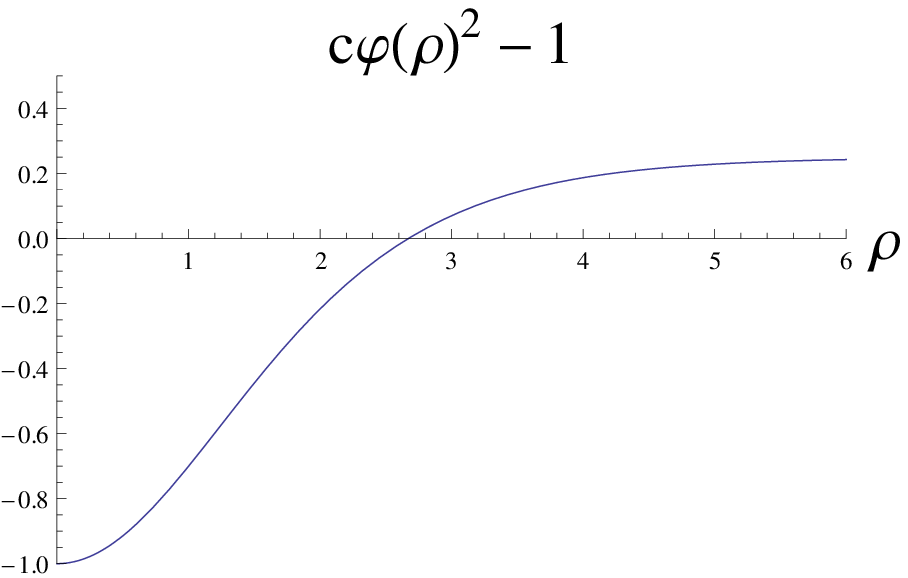} 
\includegraphics[width=7cm]{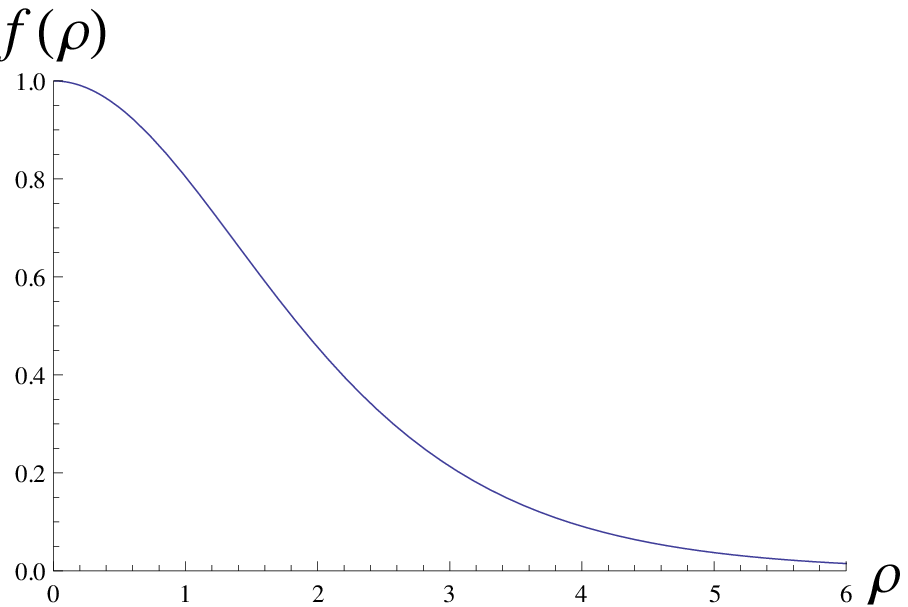}
\includegraphics[width=7cm]{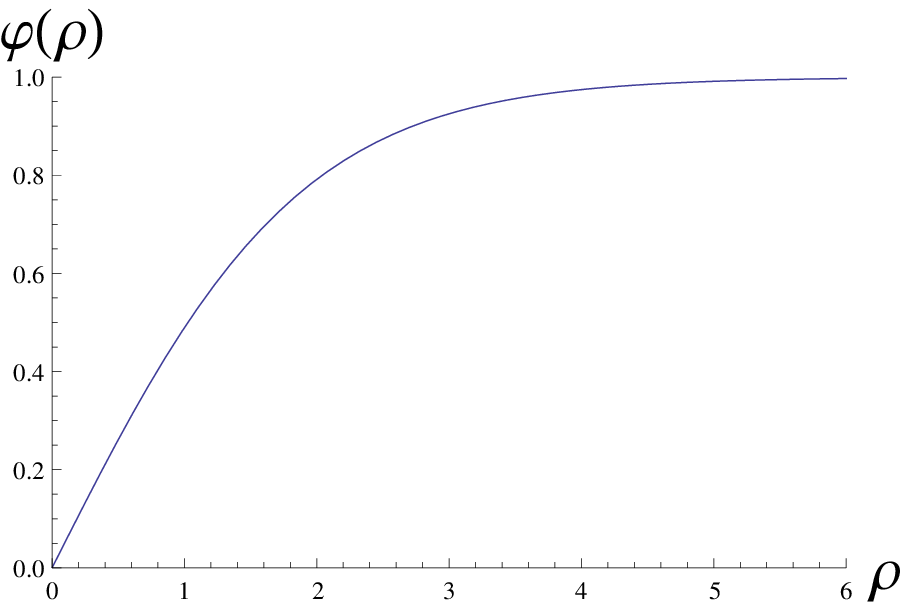}
\includegraphics[width=7cm]{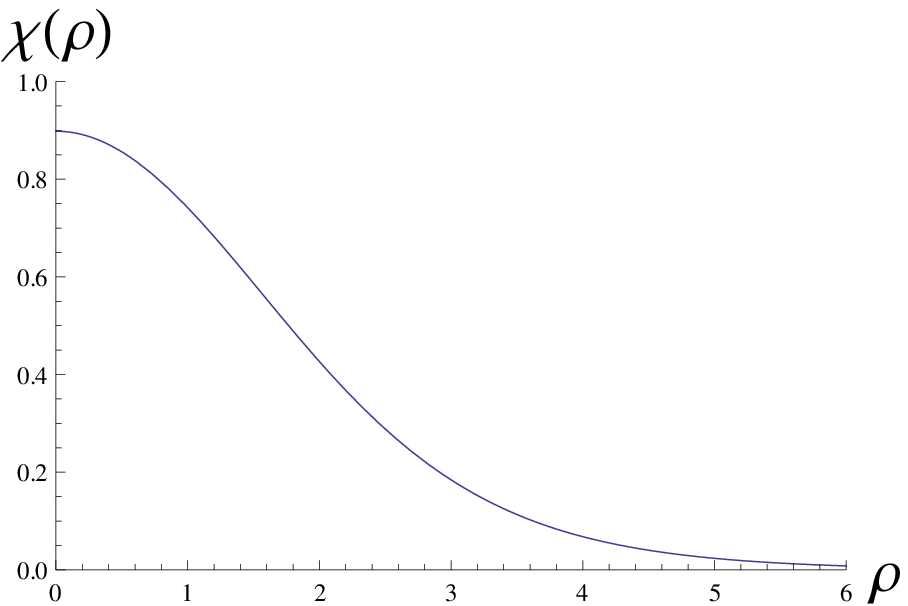}
 \caption{$b=1$, $c=1.25$, $\beta=8$}
\end{minipage}
\qquad
\begin{minipage}{7cm}
\includegraphics[width=7cm]{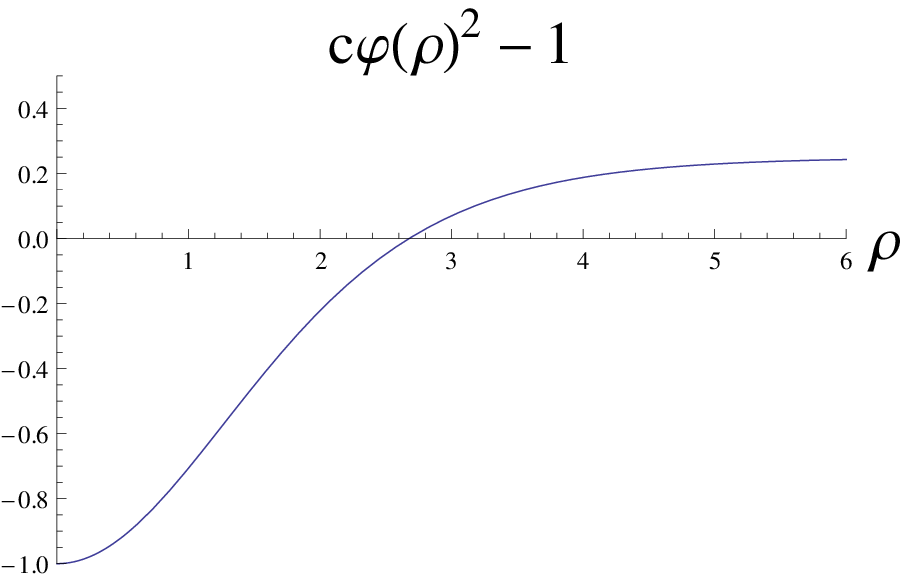} 
\includegraphics[width=7cm]{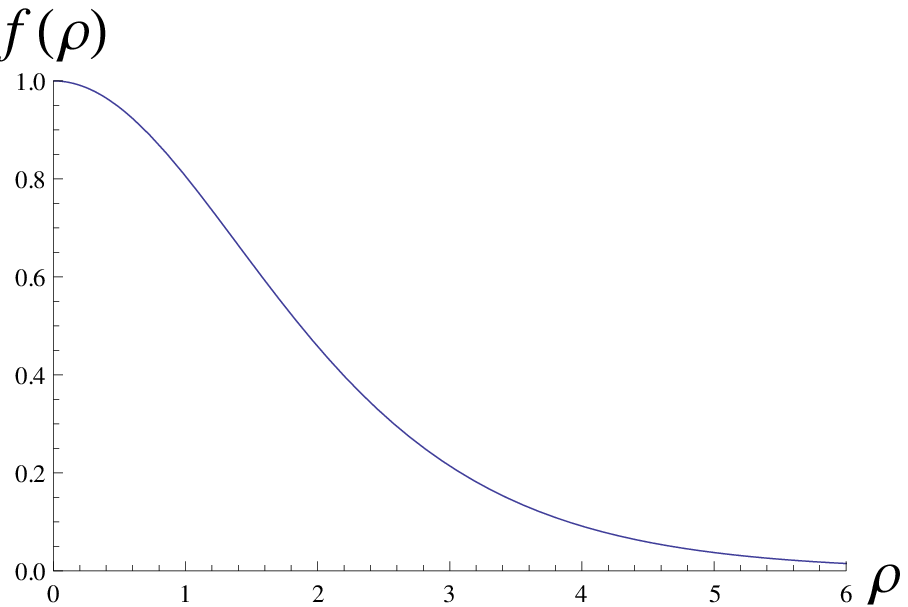}
 \includegraphics[width=7cm]{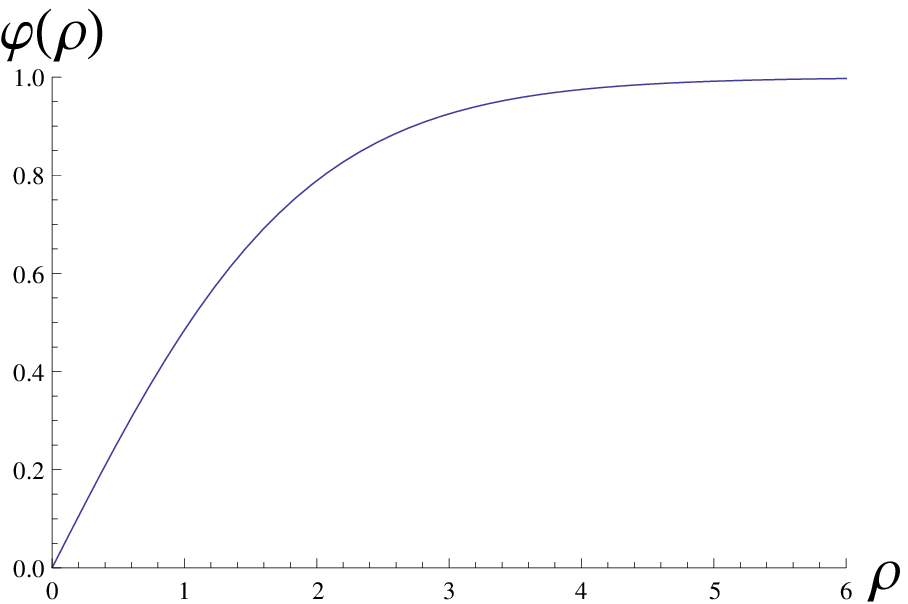}
 \includegraphics[width=7cm]{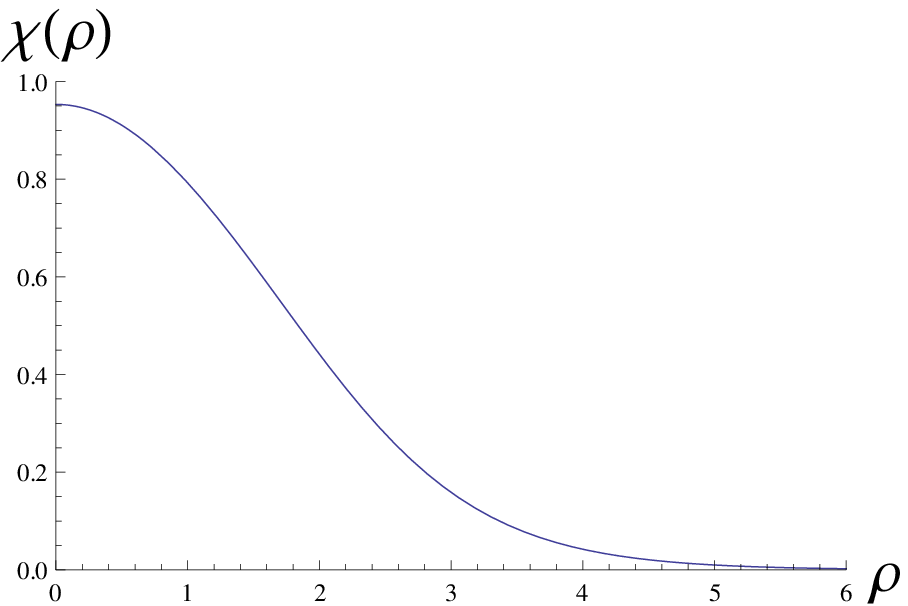}
  \caption{$b=2$, $c=1.25$, $\beta=16$}
\end{minipage}
\end{figure}

\section{Conclusions}

We discussed the theory supporting strings with
extra (rotational) moduli   
on the string world sheet. 
Our numerical analysis demonstrates that it is not difficult 
to endow the ANO string with such moduli following a strategy 
similar to that  used by Witten in constructing cosmic strings.
Our discussion was carried out  in the quasiclassical approximation.

When the bulk model is deformed by a
spin-orbit interaction  a number of entangled terms emerge on the string world sheet.
Quantum effects on the string world sheet
(which  can be made arbitrarily small with a judicious choice of parameters)
is a subject of a separate study.

\newpage

\section*{Acknowledgments}

We are grateful to Alex Kamenev, W. Vinci, and G. Volovik for inspirational  discussions.
The work of M.S. is supported in part by DOE grant DE-FG02- 94ER-40823. 
The work of A.Y. is  supported 
by  FTPI, University of Minnesota, 
by RFBR Grant No. 13-02-00042a 
and by Russian State Grant for 
Scientific Schools RSGSS-657512010.2.


\end{document}